\documentclass[12pt]{article}

\relax

\usepackage[letterpaper, margin=.85in]{geometry}

\def\keywordname{{\bfseries \emph{Keywords}}}%
\def\keywords#1{\par\addvspace\medskipamount{\rightskip=0pt plus1cm
		\def\and{\ifhmode\unskip\nobreak\fi\ $\cdot$
		}\noindent\keywordname\enspace\ignorespaces#1\par}}

\newenvironment{abstracts}[1]
{\renewcommand{\abstractname}{\large\bfseries\scshape#1}\begin{abstract}}
	{\end{abstract}}

\usepackage{natbib}

\usepackage{graphicx}
\graphicspath{{converted_graphics/}}

\usepackage{amssymb}
\usepackage{amsmath}
\usepackage{amsfonts}     

\usepackage[utf8]{inputenc} 
\usepackage[T1]{fontenc}

\usepackage{hyperref}       % hyperlinks
\usepackage{url}            % simple URL typesetting
\usepackage{doi}

\usepackage{setspace}
\usepackage{lipsum}	

\usepackage{authblk}

\usepackage{graphicx}
\usepackage{overpic}
\usepackage{natbib}
\graphicspath{{converted_graphics/}}
\usepackage[T1]{fontenc}
\usepackage{lineno}
\usepackage{setspace}
\usepackage{amssymb}
\usepackage{amsmath}
\usepackage{authblk}
\usepackage{color}
\usepackage{hyperref}
\usepackage{authblk}
\usepackage{caption}
\usepackage{subcaption}
\usepackage{lscape}

\begin{document}
\setlength{\parskip}{8pt}

\title{Maintaining human wellbeing as socio-environmental systems undergo %climate driven 
regime shifts}
\author[1]{Andrew R. Tilman}
\author[2,3]{Elisabeth H. Krueger}
\author[4]{Lisa C. McManus}
\author[5]{James R. Watson}
\affil[1]{\small USDA Forest Service, Northern Research Station}
\affil[2]{Institute for Biodiversity and Ecosystem Dynamics, University of Amsterdam}
\affil[3]{High Meadows Environmental Institute, Princeton University}
\affil[4]{Hawaii Institute of Marine Biology, University of Hawaii at Manoa}
\affil[5]{College of Earth, Ocean and Atmospheric Sciences, Oregon State University}
\date{September 8, 2023}
\maketitle

\begin{abstracts}{Abstract}

Global environmental change is pushing many socio-environmental systems towards critical thresholds, where ecological systems' states are on the precipice of tipping points and interventions are needed to navigate or avert impending transitions. Flickering, where a system vacillates between alternative stable states, is touted as a useful early warning signal of irreversible transitions to undesirable ecological regimes. However, while flickering may presage an ecological tipping point, these dynamics also pose unique challenges for human adaptation. In this work, we link an ecological model that can exhibit flickering to a model of human adaptation to a changing environment. This allows us to explore the impact of flickering on the utility of adaptive agents in a coupled socio-environmental system. We highlight the conditions under which flickering causes wellbeing to decline disproportionately, and explore how these dynamics impact the optimal timing of a transformational change that partially decouples wellbeing from environmental variability. The implications of flickering on nomadic communities in Mongolia, artisanal fisheries, and wildfire systems are explored as possible case studies. Flickering, driven in part by climate change and changes to governance systems, may already be impacting communities. We argue that governance interventions investing in adaptive capacity could blunt the negative impact of flickering that can occur as socio-environmental systems 
pass through tipping points, and therefore contribute to the sustainability of these systems.
\end{abstracts}
\onehalfspacing

\section*{Introduction}
%\linenumbers

Global change impacts, including those resulting from climate change and socioeconomic transitions, are altering the environment and threatening human livelihoods. For example, worsening drought conditions lead to an increased risk of wildfires~\citep{mckenzie2004climatic}, and global mean sea-level rise  threatens the habitability of low-elevation coastal zones~\citep{vitousek2017doubling}. In ecological systems, these environmental changes are linked to phenomena known as regime shifts, wherein there is a ``large persistent change in the structure and function of an ecosystem'' \citep{biggs2012}. Prominent case studies include shifts from productive coral-dominated reefs to degraded systems dominated by macroalgae \citep{mumby2007}, and shifts from a highly vegetated to a barren state in arid landscapes \citep{rietkerk2004}. Because these transitions have implications for the ability for humans to thrive in these systems, much attention has been focused on identifying indicators to serve as `early warning signals' for impending catastrophic changes~\citep{scheffer2009early,bauch2016early}. Whether informed by early warning signals or not, people experiencing ecological regime shifts can attempt to adapt to these changing conditions to maintain their wellbeing. These adaptation measures could include minor changes to practices such as switching target species in fisheries~\citep{katsukawa2003simulated}, or more major changes such as migration to pursue an alternative livelihood elsewhere~\citep{adamo2010environmental}.

While identifying and attempting to avoid a transition to an undesirable alternative state remains a challenge, in this paper we focus on how people navigate ecological regime shifts.  Adapting to environmental change while maintaining wellbeing poses unique challenges, especially in noisy systems. The global extent of human societies demonstrates the ability of people to adapt to (and prosper under) a broad range of ecological regimes. Other organisms also demonstrate this adaptive capacity in their evolutionary response to environmental change~\citep{carlson2014evolutionary}. However, in social and environmental systems, stochasticity can result in a phenomenon known as ``flickering'' when the system approaches a regime shift~\citep{taylor1993flickering, wang2012flickering, gatfaoui2019flickering}. Flickering describes how a system switches between alternative stable states as a result of stochasticity. In the context of an impending regime shift, flickering leads to periods of time which resemble the status quo alternating with times defined by a novel socio-environmental state. This presents a unique challenge for adaptive agents: which regime should one adapt to and when should one shift practices to align with the expected post-regime-shift environment? There may be cases where agents themselves flicker between alternative livelihoods in an attempt to adapt to the intermittent shifts in environmental regimes. Our results suggest that in systems where people have limited adaptive capacity, flickering can yield marked declines in utility.

\begin{figure}
    \centering
    \includegraphics[width=.8\textwidth]{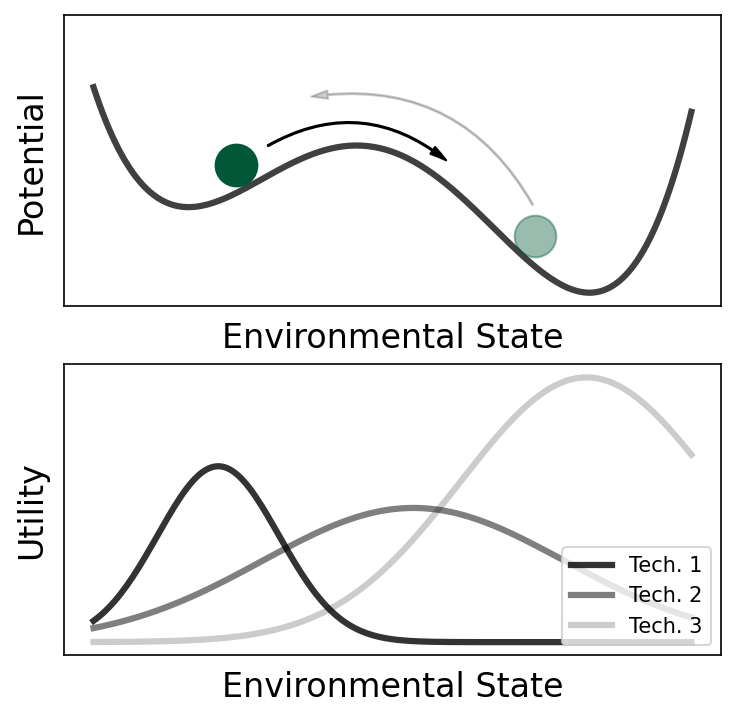}
    \caption{Illustrations showing how alternative stable environmental states can be conceptualized as basins where a ball subject to stochastic perturbations (shown by arrows) may settle, as illustrated in the first panel. Noise may cause the environmental state to tip from a low to a high potential state or vice versa, a process termed flickering. In agricultural systems there are often a range of technologies / approaches / strategies that have environmentally dependent utility curves (3 are illustrated in the lower panel). As environmental states shift, the most favorable strategy changes. In this paper, we consider the case where a continuum of environmentally dependent utility curves exist and individuals adapt to environmental change by shifting the peak of their utility curve in an attempt to track changing environmental conditions. We show that environmental regimes with flickering pose unique challenges for adaptive agents and can lead to troughs in average utility.}
    \label{fig:concept}
\end{figure}

The importance of adaptation to flickering and nonlinear ecological dynamics can be illustrated through environmentally dependent utility functions. Such utility functions can arise when they depend on environmental productions functions. These functions provide a mapping between some measure of the environmental state and production or output. They are typically approximated by hump-shaped functions~\citep{schlenker2009nonlinear} (see Figure 1b). In agriculture, for example, a production function for a particular crop may be dependent on temperature. Multiple functions with peaks that span different ranges of the environmental state (along the x-axis) represent production under different strategies. In this context, climate adaptation can be thought of as people shifting their production strategy, and thereby transitioning across the different production functions in order to maintain high levels of output despite environmental change. In agriculture, this could be achieved by choosing different varieties of crops as temperature increases.

However, these production functions do not explicitly account for the non-linear dynamics associated with coupled social-ecological systems. Whereas average temperature in a region may slowly increase in response to climate change and lead to the expectation of a steady advance through a series of production strategies, underlying environmental conditions that shape productivity may exhibit far more complex dynamics in response to gradual global change. To integrate these effects, we consider an environmental model that has the potential for nonlinear dynamics in response to gradual change in an underlying parameter. These nonlinear dynamics associated with changes in the environmental state are often depicted using a well-potential diagram (Figure 1a). Well-potential diagrams can illustrate how ecosystems (and social-ecological systems) exhibit alternative stable states, and how the resilience of a particular state may be eroded by a relatively slow-changing parameter like average temperature. As the resilience of one basin of attraction is diminished and an alternative basin arises, flickering can occur prior to a tipping point being crossed. After the tipping point is crossed the old basin ceases to exist and the system transitions to the alternative stable state.

In contrast to production functions that are dependent on a gradually changing environmental parameter, the utility functions we model depend on an environment with complex dynamics and can lead to highly stochastic utility as the environment flickers between alternative stable states and as people struggle to adapt to this volatility. Additionally, due to the hump-shaped relationship between the environment and productivity, environmental variability will tend to depress average productivity due to non-linear averaging. By integrating the non-linear dynamics of coupled social-ecological systems -- especially the dynamical flickering associated with some regime shifts -- with the economic concept of environmentally-dependent production functions, we provide new insight into the potential impacts of regime shifts on human wellbeing. We explore the impact of people's adaptive capacity on their ability to track environmental change and maintain their wellbeing. We also examine when transformational change to novel strategies which buffer individuals against environmental change should be adopted given flickering dynamics.
 
We develop a mathematical model to describe the impact of alternative stable states and flickering on the utility of adaptive agents in a coupled socio-environmental system. We use this model to highlight the conditions wherein flickering has the largest negative impact on people's wellbeing and explore how the timing of people's transitions to new strategies should relate to the timing to environmental transitions caused by flickering and tipping points.  We discuss several possible case studies that illustrate how these dynamics could or may be impacting communities, and primarily contextualize our model based on the response of Mongolian nomadic pastoralists to global change. Nomadic pastoralist communities have been among the hardest hit by the consequences of global change. New political regimes have shifted borders, fundamental changes to economic systems have limited the prospects of nomadic ways of life, and extended drought periods have put the health of livestock herds at risk. Using available data from the literature and other publicly available sources, we discuss how global change impacted these communities, with long-term effects that include the migration of people away from their traditional homes. We also discuss how flickering dynamics could have similar consequences for artisanal fishing communities impacted by marine climate shocks and communities impacted by wildfires.  

\section*{Socio-environmental model} 
The model has two components: an ecological component where nonlinear dynamics (and in particular flickering) occur, and a social component, where agents adapt their production frontier to align with the environment and maximize their utility.

\subsection*{Ecological dynamics}
In our model, we assume that there is an ecological state, $x$, that can be described by logistic growth and experiences sigmoidal harvest rate $c$ following a type-3 functional response~\citep{holling1959components}. This approach forms the basis for well-studied ecological models that can exhibit alternative stable states and hysteresis~\citep{may1977,scheffer1989}; models of this form have also been used to study flickering dynamics~\citep{dakos2012methods}. The ecological state, $x$, could represent the abundance of forage plants in the context of grazing systems, fish biomass in the case of fisheries, or forest biomass in the case of wildfires. The discrete-time stochastic dynamics of $x$ are described by

\begin{equation}
    x_{t+1} = \left(r x_t\left(1-\frac{x_t}{K}\right) - c\frac{x_t^2}{x_t^2+h^2}\right) + \left(1 + i_t\right)x_t
\end{equation}
where $r$ is intrinsic growth rate of $x$, $K$ is its carrying capacity, $h$ is the half-saturation constant (i.e., the resource level at which half of the maximum extraction rate is reached), and $i_t$ a noise term that models environmental shocks. We assume that $i_t$ is time-correlated red noise governed by

\begin{equation}
    i_{t+1} = \left(\left(1-\frac{1}{T}\right)i_t +  \eta_t\right),
\end{equation}
where $i_t x_t$ is the magnitude of the stochastic reduction or increase of the resource at time $t$, $T$ is the time scale over which noise becomes uncorrelated, and $\eta_t\sim \mathcal{N}(0,\beta^2)$ is an element of a series of independent identically distributed normal error terms.

\begin{figure}
    \centering
    \includegraphics[width=\textwidth]{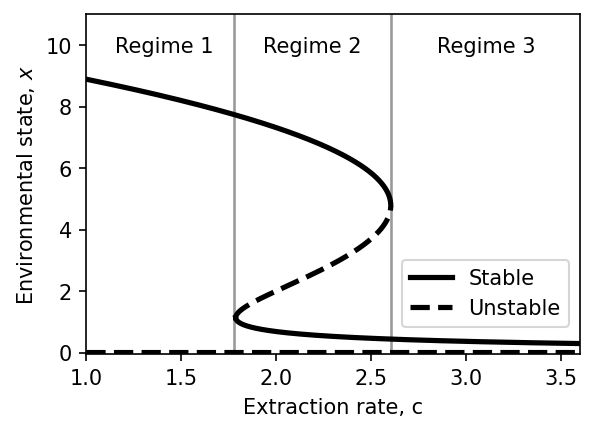}
    \caption{For low and high extraction rates, the system has only one stable equilibrium. For intermediate extraction rates, bistability occurs and the potential for flickering dynamics arises. We use this distinction to classify our system into three distinct dynamical regimes. In regime 1, only the high environmental state is stable, regime 2 exhibits bistability and potentially flickering dynamics, and only the low environmental state is stable in regime 3. For any extraction rate, actual dynamics of the environment will fluctuate about their equilibria due to stochasticity.}
    \label{fig:basins}
\end{figure}

In the absence of noise (i.e., when $\beta=0$), this system can exhibit a range of dynamics. For large values of $r$, discrete-time logistic systems such as this can exhibit cyclic or chaotic dynamics~\citep{may1974biological}. To simplify our analyses, we restrict our attention to those cases where cyclic and chaotic dynamics do not occur in the absence of noise. For harvesting rates $c$, that are low, the system has a single stable equilibrium corresponding to an environmental state of abundance. For high values of $c$, the sole stable equilibrium is a depleted environmental state. For intermediate values of $c$ there exists a region with multiple stable equilibria. In this intermediate regime, the inclusion of noise can lead to a flickering dynamic where the system makes irregular jumps from the high to low environmental basins of attraction~\citep{dakos2012methods}. Figure~\ref{fig:basins} shows the stable and unstable equilibrium states across a range of harvesting values, and illustrates that for intermediate extraction rates, the system has two alternative stable states. 

\subsection*{Human adaptation and wellbeing}
To model human wellbeing in response to a constantly changing environment, we assume that people can adapt their practices to be in alignment with the environmental state, but that this adaptation process takes time. The rate at which individuals can adapt to a changing environment depends on their adaptive capacity. Here, we conceptualize adaptation as individual or collective actions that allow individuals to be as successful as possible, given the current state of the environment. Adaptation allows individuals to shift the peak of their production function so that it aligns with the current environmental state. Potential avenues for adaptation are myriad and depend on the context of the case study. For pastoralist systems, they include moving to better locations when the local resource level is low, implementing irrigation systems to bolster productivity, and storing feed for cattle~\citep{chen2015}. In agriculture, it could include adjusting the timing of planting and harvesting, and changing (or diversifying) the varieties of crops grown.   

We employ a very simple model that captures the notion of adaptive capacity and allows us to explore the consequences of decreased adaptive capacity on people's wellbeing. We let $y_t$ be the environmental state to which individuals are most well adapted. When $y_t=x_t$, agents achieve the highest possible payoff given the current environmental state. The adaptation of individuals to environmental change is constrained by their adaptive capacity, $l$. When $l=1$, individuals adapt fully in one time step to the current environmental state. On the other hand, when $l<<1$, it will take much more time for adaptation to a particular environmental state to be achieved. We model the dynamic of adaptation as a deterministic discrete-time dynamical system governed by

\begin{equation}
    y_{t+1} = l(x_t-y_t) + y_t.
\end{equation}
 
 We construct a utility function that depends on the best possible payoff, $\pi(x)$, that can be attained given the current environmental state, $x$, and on the extent to which there is a divergence between the environmental state and the state to which individuals are most well adapted. We assume that the highest achievable payoff given an environmental state $x$ is a linearly increasing function $\pi(x)$, such that the potential for high payoffs improves with the state of the environment (Figure 3). We let this payoff be the utility that is achieved by an individual who is perfectly adapted to state $x$, (i.e. $y(t)=x(t)$).
When there is some degree of divergence between the current environmental state and an individual's adaptation, then utility decreases. We assume that utility can be described as a Gaussian function of environmental adaptation, $y$, given by

\begin{equation}
    U(x,y) = \pi(x)\;\exp\left(\frac{-\ln(2)(x-y)^2}{a^2}\right)
\end{equation}

where $a$ defines the degree of misadaptation at which utility is cut in half from its peak. This utility function adheres to our assumption that when environmental adaptation, $y$, is equal to the current environmental state, $x$, then utility, $U(x,y)$, is equal to the payoff $\pi(x)$.

\begin{figure}
    \centering
    \includegraphics[width=\textwidth]{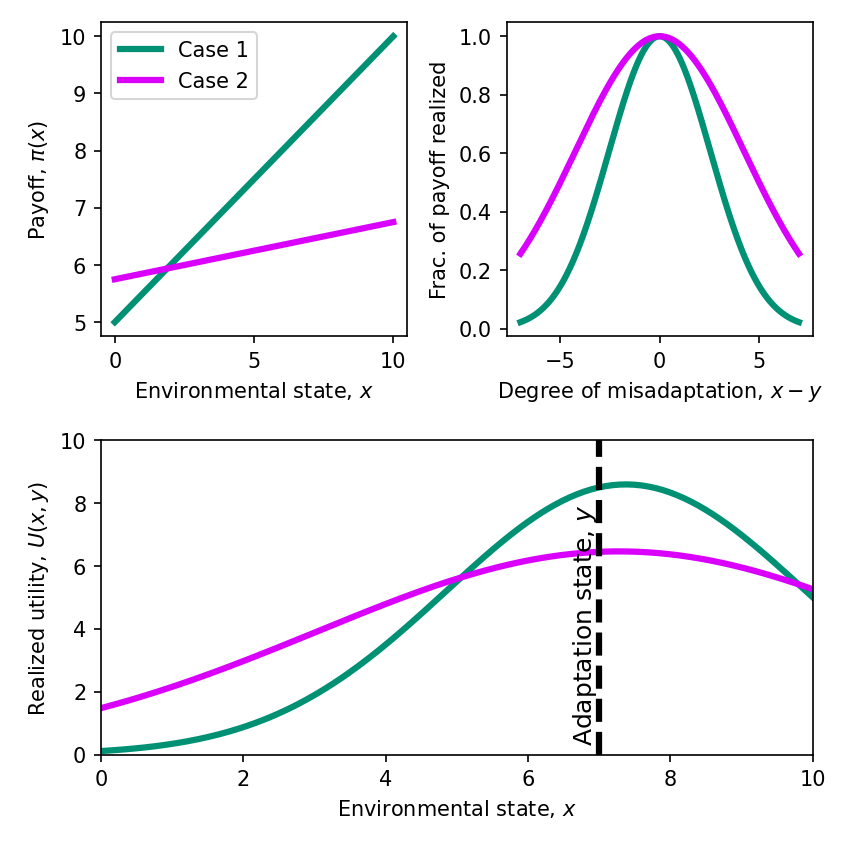}
    \caption{Illustrations showing two representative cases. In the first panel, the relationship between the environmental state and maximum payoff, $\pi(x)$ is shown for a case where the payoffs are highly sensitive to the environmental state (Case 1) and where payoffs vary less in response to different environmental states (Case 2). The second panel shows the relationship between the misadaptation to the environment and how much of payoffs individuals realize as utility. In Case 1, individuals are more sensitive to misadaptation than Case 2. Finally, in the bottom panel, these impacts are aggregated and utility is shown as a function of the state most well adapted to ($y$) for a current environmental state of $x=7$.}
    \label{fig:utiladapt}
\end{figure}

\subsection*{Human-environmental dynamics}
The coupled dynamics of the environment and adaptation can be described by the system of difference equations 

\begin{align}
    x_{t+1} &= \left(r x_t\left(1-\frac{x_t}{K}\right) - \frac{cx_t^2}{x_t^2+h^2}\right) + \left(1 + i_t\right)x_t\\
    i_{t+1} &= \left(\left(1-\frac{1}{T}\right)i_t +  \eta_t\right)\\
    y_{t+1} &= l(x_t-y_t) + y_t,
\end{align}
where $i$ is a red noise term, $x$ is the state of the environment, and $y$ is the environment to which individuals are most well adapted. Figure~\ref{fig:basins} shows that for low extraction rates, the sole stable equilibrium of the system is a high environmental state, but as extraction increases, a tipping point is crossed and the environmental state collapses. Layered on top of this tipping point is human adaptation, which influences wellbeing. We start by focusing solely on the dynamics of the environment and adaptation. Later we will turn our attention to the implications for wellbeing of these dynamics.

\subsection*{Results}
\subsubsection*{Environmental adaptation in three regimes}
Figure~\ref{fig:basins} identifies three regimes which exhibit qualitatively distinct environmental dynamics. In regime 1, there is a single stable equilibrium with high resource biomass. In regime 2, there are alternative stable states, one with high biomass and one with a degraded environmental state. Lastly, in regime 3, only the degraded environmental equilibrium remains. 

\begin{figure}
     \centering
     \begin{subfigure}[b]{0.45\textwidth}
         \centering
         \includegraphics[width=\textwidth]{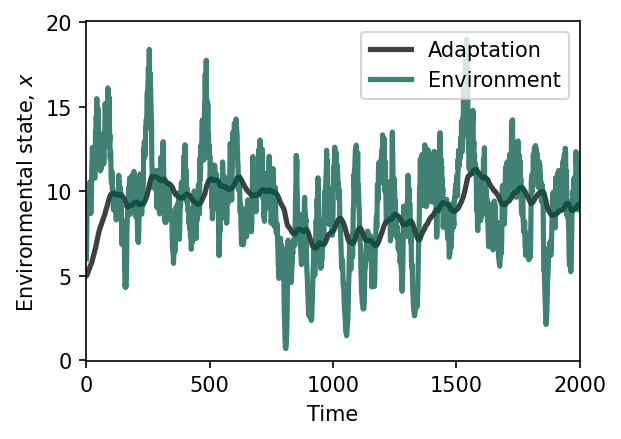}
         \caption{Regime 1: $c=1$ results in a high, but variable environment state}
         \label{fig:c_low}
     \end{subfigure}
     \begin{subfigure}[b]{0.45\textwidth}
         \centering
         \includegraphics[width=\textwidth]{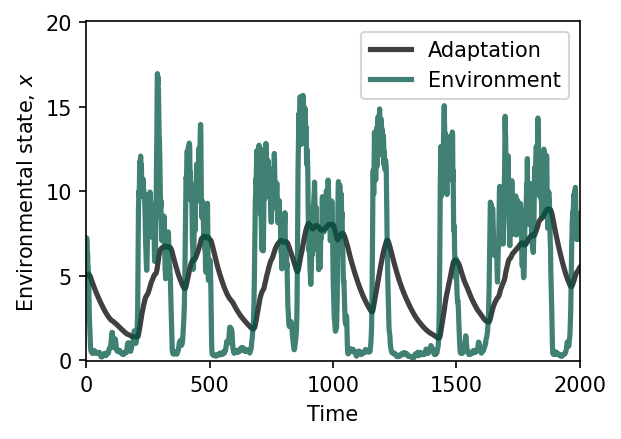}
         \caption{Regime 2: $c=1.95 $ shows flickering dynamics with brief periods of collapse.}
         \label{fig:c_mid}
     \end{subfigure}\\
          \begin{subfigure}[b]{0.45\textwidth}
         \centering
         \includegraphics[width=\textwidth]{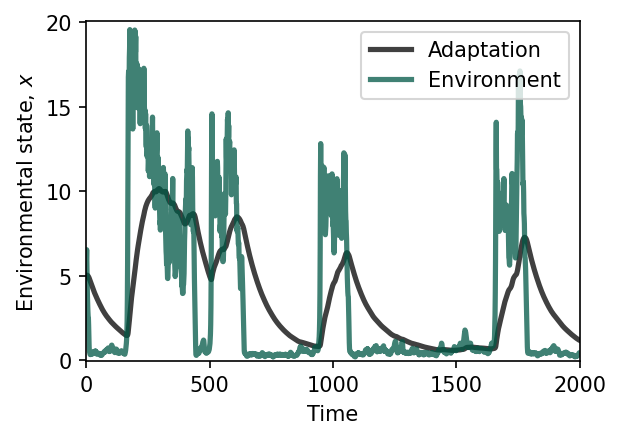}
         \caption{Regime 2: $c=2.45$ results in flickering dynamics with longer periods in a collapsed state.}
         \label{fig:c_mid2}
     \end{subfigure}
     \begin{subfigure}[b]{0.45\textwidth}
         \centering
         \includegraphics[width=\textwidth]{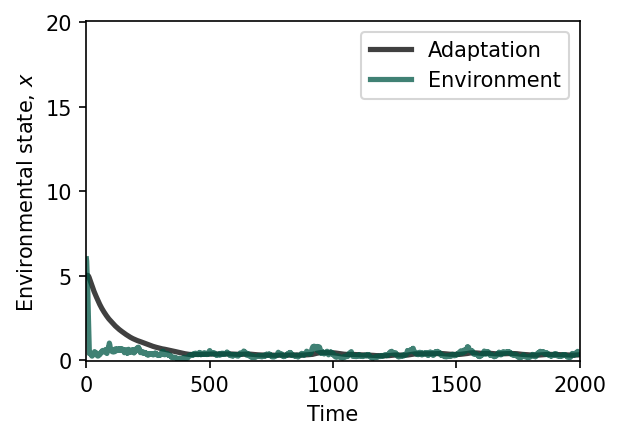}
         \caption{Regime 3: $c=3.1$ yields system collapse, with the possibility of brief periods which resemble recovery.}
         \label{fig:c_high}
     \end{subfigure}
        \caption{Temporal dynamics of the environment and adaptation across three regimes. When environmental dynamics fall into regime 1 or 3, individuals adaption generally falls within the range of day-to-day environmental variability. For flickering environmental dynamics as seen in regime 2, when the environment flips from one basin of attraction to another, there are extended periods where individuals are significantly misadapted to the environment. This has important implications for wellbeing.}
        \label{fig:three sims}
\end{figure}

We are motivated by the scenario in which historical conditions of the system correspond to regime 1. In other words, we start in a scenario where high biomass predominates. Nevertheless, the system is stochastic, and fluctuation about this high-biomass equilibrium can be significant in magnitude. The dynamics of adaptation are governed by the same equation across all three regimes, with agents adjusting their practices toward the current environmental state. Figure~\ref{fig:c_low} shows that dynamics of the environment in regime 1 exhibit significant variation but that adaptation generally falls within the range of environmental variability. 

In response to shifting economic structures and  environmental change, we assume that the extraction rate, $c$, in the system will increase over time and the system will eventually fall within regime 2. Figure~\ref{fig:c_mid} and Figure~\ref{fig:c_mid2} with $c$ values corresponding to regime 2 show examples of flickering environmental dynamics. Whereas adaptation largely remains within the range of environmental variability in regime 1, in regime 2 when the system flips from one equilibrium region to the other, there are significant time periods where adaptation is significantly misaligned from the state of the environment. We show that this has important implications for the utility (wellbeing) of agents in regime 2.

Figure~\ref{fig:c_high} shows a case where the extraction rate is high enough that the system has been pushed beyond a tipping point and is in regime 3. The only stable equilibrium is a collapsed environmental state, but stochastic dynamics may nonetheless lead to ephemeral periods where environmental dynamics resemble historical conditions. In regime 3, agents are more or less able to maintain close adaptation to the environment. Because of this, we expect utility to approach the maximum payoffs that can be attained under perfect adaptation.

\subsubsection*{Wellbeing and environmental regimes}
Wellbeing depends both on the maximum profitability that could be achieved given environmental conditions, as well as the degree to which agents are adapted to the environmental state. In this section, we examine how wellbeing (i.e., utility) depends on the environmental extraction rate, $c$. 

\begin{figure}
    \centering
    \includegraphics[width = .9\textwidth]{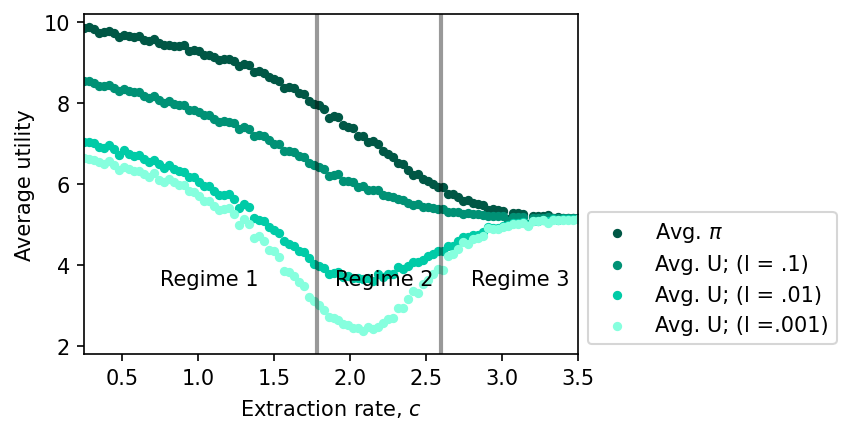}
    \caption{Maximum attainable payoff given perfect environmental adaptation and realized utility under actual environmental adaptation for a range of extraction rates, $c$ and levels of adaptive capacity, $l$. Flickering can occur in regime 2 for intermediate extraction rates. Flickering dynamics present unique challenges for adaptive agents, especially those with limited adaptive capacity. When agents have low adaptive capacity, flickering leads to a utility trough that does not occur for agents with high adaptive capacity.}
    \label{WellbeingFlickering}
\end{figure}

As discussed, extraction rates structure the system into three qualitatively distinct regimes. Figure ~\ref{WellbeingFlickering} shows how average payoff assuming perfect adaptation and average utility depend on which regime the extraction rate falls within and on the level of adaptive capacity, $l$. Figure~\ref{WellbeingFlickering} shows the maximum average payoff, 

\begin{equation}
\overline{\pi} = \frac{1}{t_\text{max}}\sum_{t=1}^{t_\text{max}}\pi\left(x_t\right),
\end{equation}
that could be achieved through time for different fixed rates, $c$, of environmental extraction.  The figure also represents average utility,

\begin{equation}
\overline{U} = \frac{1}{t_\text{max}}\sum_{t=1}^{t_\text{max}} U\left(x_t,y_t\right),
\end{equation}
for several levels of adaptive capacity, $l$. Unlike $\overline\pi$, $\overline U$ depends on both the state of the environment, $x_t$ and adaptation, $y_t$. For higher values of adaptive capacity, $l$, the qualitative pattern of average utility mirrors that of average payoff. When $l=0.1$, Both average payoff, $\overline\pi$ and average utility, $\overline U$ gradually decline as the extraction rate increases through the three regimes.

For moderate to low levels of adaptive capacity, (e.g. $l=0.01$ to $l=0.001$) the qualitative patterns seen in average payoff, and in average utility under high adaptive capacity no longer hold. In these cases, the flickering dynamics of regime 2 exacts a costly toll on average utility. The repeated switching between high and low environmental states leads to extended periods of misadaptation that diminish average utility and creates a noticeable utility trough in regime 2.  

In regime 3, the collapsed environment doesn't vary as widely, and even agents with limited adaptive capacity can eventually adapt their behaviors to the permanently degraded environmental state. This leads to an increase in the average utility of agents with low adaptive capacity and a convergence of utilities and payoffs across all scenarios. This result raises important questions about the usefulness of flickering as an early warning signal in socio-environmental systems. Flickering dynamics can depress average wellbeing by creating highly variable environmental conditions with abrupt shifts that require significant time to adapt to, especially for agents with low adaptive capacity. 

Our simulations suggest that rather than being an early warning signal, in socio-environmental systems, flickering can be a uniquely challenging regime for agents with low adaptive capacity.

\subsubsection*{Transformational change} 
In this section we consider the option of transformational change, where agents can choose once to dramatically overhaul their practices and adopt a more generalist approach. 

In Figure~\ref{fig:utiladapt} we showed two alternative cases for the structure of the payoff and utility functions. The first case, where payoffs are highly sensitive to the environmental state and utility is sensitive to misadaptation, has been the focus of the preceding simulation results. In essence, we have assumed that individuals attempt to become specialists in their environment. The second case shown in Figure~\ref{fig:utiladapt} highlights an alternative possibility, that payoffs are less sensitive to environmental conditions and utility is less sensitive to misadaptation. This corresponds more closely with a generalist strategy, where peak payoffs may never be as high, but adaptation to precise environmental conditions is less important. 

\begin{figure}
    \centering
    \includegraphics[width = .9\textwidth]{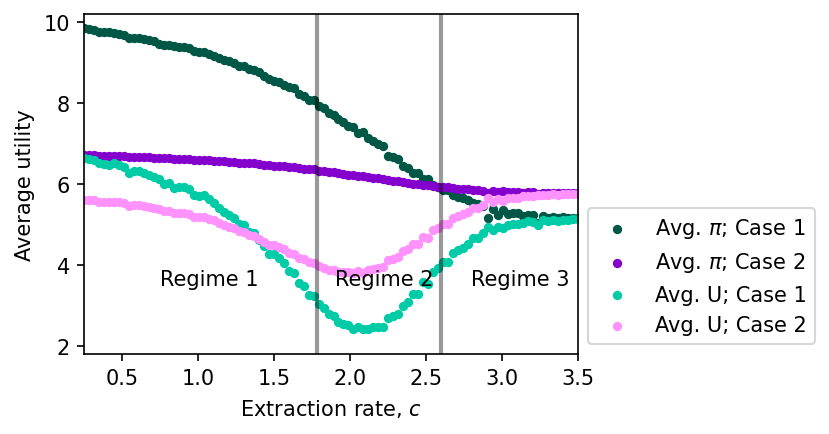}
    \caption{If agents are posed with the problem of when to transform their practices from a specialist approach (Case 1) to a generalist approach (Case 2) in response to a steadily increasing extraction rate in the system, the answer will depend on whether they take adaptation into account. Case 1 corresponds to the status quo, where payoffs are highly dependent on the state of the environment and highly sensitive to misadaptation. Case 2 corresponds to a transformative change, where payoffs are less sensitive to the state of the environment and misadaptation. Case 1 resembles a setting where individuals specialize their practices adapted to environmental conditions. Case 2 represents a transformative change where individuals' wellbeing is less sensitive to the environmental and adaptation states. As the extraction rate increases, agents are eventually better off adopting transformative change. However, the timing of this shift depends on the adaptive capacity of agents. If people are always perfectly adapted to the environment, then payoffs under the status quo remain higher than those under transformation until after the system passes its tipping point at the cusp of regimes 2 and 3. On the other hand, when people are limited in their adaptive capacity, the optimal transformation timing is earlier, as illustrated by the intersection of the pink and green points (Utility curves) in regime 1. In this case, people are better off transforming before the system even reaches the flickering regime.}
    \label{Wellbeingadaptvstrans}
\end{figure}

Transformation to a generalist approach is an alternative to specialist adaptation to one's environment. Under transformation, agents fundamentally alter their practices in order to switch their payoff and utility functions from those in case 1 to those described by case 2. Figure~\ref{Wellbeingadaptvstrans} shows the maximum average payoff, $\overline \pi$, of a specialist approach (case 1, in dark green) and a transformational generalist approach (case 2, in dark purple). In this example, the payoff functions are such that if agents were always perfectly adapted to the environment, they would choose the transformational approach as the system nears regime 3 and the dark purple circles rise above the dark green circles. However, agents will not always be perfectly adapted to the current environment, so their realized average utility will fall below the maximum attainable average payoff. In this case, agents would increase their average utility by adopting the transformational approach (case 2) while the system is still in regime 1. After transformation, the flickering dynamics experienced in regime 2 are far less detrimental to average utility. After crossing into regime 3, the average utility of the transformation approach remains higher than the baseline specialist approach. Other payoff and utility functions can be constructed where the transformational generalist approach is only favored during the flickering regime. In this case, analyzing average payoffs could indicate that transformation is never beneficial, while an analysis focusing on realized average utility could show that transformation can help agents navigate flickering critical transitions without suffering as greatly from the utility trough that might otherwise occur.

\begin{table}[ht]
\begin{center}
    \begin{tabular}{c|c|l } 
     Variable & Range of values & Description \\ 
     \hline
     $x_t$ & 0--20 & Current environmental state  \\
     $y_t$ & 0--20 & Current adaptation state \\
     $l$ & .001-1 & Adaptation rate\\
     $i_t$ & & Auto-correlated red noise \\
     $T$ & 30 & Timescale over which noise becomes uncorrelated \\
     $\eta_t$ & 0&  i.i.d. normal error term \\
     $\beta$ & .07 & The standard deviation of $\eta$'s \\
     $r$ & 1  & Resource growth rate\\
     $K$ & 10 & Resource carrying capacity\\
     $c$ & 0--4  & Extraction rate \\
     $h$ & 1 & Extraction half-saturation constant \\ 
     ${\pi}(x)$ &5--10& Environmentally dependent payoffs \\
     $U(x,y)$ & 0--$\pi(x)$ & Utility as a function of environmental and adaptation states\\
     $a$ & 3-5 & Value of $|x-y|$ for which $U(x,y)=1/2\;\pi(x)$ \\

    \end{tabular}
     \caption {Variables and parameters in the model, their approximate range of values, and meanings. Exact parameter values used for each figure available in Table~\ref{SI Table: paramVals}.}
\end{center}
\end{table}
\section*{Case studies}
\subsection*{Nomadic pastoral systems in Mongolia}

Climate change, land degradation and socio-political disruptions threaten the livelihoods of tens of millions of nomadic pastoralists who inhabit the  semi-arid areas of Central Asia, the Middle East, North Africa and the Sahel zone. Nomadic pastoralists move seasonally between several locations that provide adequate conditions
for livestock grazing and water access \citep{sneath2003}.

Mongolia is the worlds most sparsely populated country, has an arid to semi-arid climate, few forests, and very limited arable land~\citep{MongoliaStats}. 
Nomadic lifestyles are well adapted to these highly variable landscapes and climate conditions. In Mongolia, nomadic pastoralists employ many adaptive strategies including storage of fodder for use during poor grazing conditions, mobility to seek out better rangelands, and communal pooling of resources and labor \citep{fernandez2015}. Today, about one third of the country’s population continues to be nomadic or semi-nomadic, while two thirds live in urban areas (See figure~\ref{SI fig: PopulationUrban}).

Over the past 70 years, Mongolia has experienced an increase in average annual temperatures of 2.1°C \citep{lkhagvadorj2013}, which has been accompanied by the proliferation of extreme weather conditions, in particular since the mid-1990 \citep{Enebish2020}. 
Severe winter weather events called Dzud, characterized by inaccessible grazing resources, have occurred repeatedly and with devastating effects for nomadic livelihoods, killing 33, 23 and 10 million heads of cattle in 1999, 2003 and 2010, respectively \citep{Rao2015}.

During this time of dramatic environmental change, Mongolia also experienced a political transition from a socialist to a capitalist regime. This brought about changes to the governance, institutional, and public service structures upon which both settled and nomadic people rely. Prior to the emergence of the capitalist regime in 1990, herders were organized into modular collectives of families around community centers known as \emph{`bag'}. These centers provided technical, social, health and veterinary services, as well as emergency stocks of fodder to buffer against Dzud events \citep{fernandez2015}. 
These centers bolstered the resilience of Mongolian pastoralist communities, even to severe shocks such as the loss of a family's  entire herd.  They represented type of community insurance that allowed families who lost their herd to restock the following year \citep{fernandez2015,Ahearn2018}. 

Upon the institution of a capitalist regime, these governance structures were abandoned. Bag centers and their social services all but disappeared \citep{fernandez2015}. Consequently, pastoralist families had to adapt by becoming increasingly self-reliant; they grew their livestock herds to raise income during favorable years and to be able to buffer losses during Dzud years. Livestock numbers increased from around 20-25 million during the socialist era, when livestock numbers were strongly regulated, to around 70 million today \citep{MongoliaStats}, with grazing pressure growing accordingly (See Supplementary Information, Figure~\ref{SI fig:grazing rate}). 

In response to these dynamics and the devastating Dzud losses of the past decades, external actors, such as the Asian Development Bank, pushed the Mongolian government to privatize land ownership, in order to give exclusive land use rights to individual families and reduce overgrazing \citep{sneath2003}. However, land privatization came at the cost of mobility, an essential element of pastoralism. Mobility is an adaptation to the low average productivity and high spatio-temporal variability of Mongolian rangelands~\citep{sneath2003}. Land privatization and reduced mobility led to a further deterioration of local pastures, adding to the vulnerability of the pastoralist families \citep{fernandez2015}, \citep{wang2013}. Thus, a number of compounding factors led to maladaptive measures and the loss of adaptive capacity and wellbeing: The response to extreme weather conditions and changing governance systems resulted in increased grazing pressure and may have provoked signals of environmental flickering leading to massive losses in livestock, while measures promoted by external actors aimed at tackling the loss of previous governance systems on the one hand, and the problem of overgrazing on the other hand, proved to be maladaptive.

After the devastating Dzud of 2009/10, a number of donor-initiated community-based natural resource management organizations were initiated with the aim to establish community structures that resembled those that existed during the socialist regime, which had provided adaptive capacity at the community level. Others responded to the increasingly difficult conditions for maintaining rural lifestyles by migrating to urban centers. While the rural population has remained more or less stable over the past thirty years, the urban population has doubled~\citep{MongoliaStats}(See Supplementary Information, Figure~\ref{SI fig: Population}). For some, moving to an urban setting presented new opportunities for education and employment. For others, the transformation posed insurmountable challenges. Rapid urbanization has overwhelmed urban governance institutions, and the majority of urban migrants in Mongolia have inadequate access to basic services, such as water, sanitation and electricity \citep{Terbish2016}, and are unable to support their livelihoods once livestock herds are lost to Dzud or sold during the rural-urban migration process.

For those that have remained in a (semi-) nomadic lifestyle, there is evidence of adaption to the highly variable (and potentially flickering) dynamics of weather conditions and pasture quality through new forms of communal governance \citep{fernandez2015}, and private insurance \citep{Ahearn2018}. Under these highly unstable conditions, pastoralists have to invest into costly adaptation measures. 
A study examining adaptation strategies to climate impacts in Mongolia found more than 50 adaptation strategies applied in different combinations and with different frequencies \citep{wang2013}.

While land degradation has even been reversed in some areas of the Mongolian Plateau \citep{Guo2021}, many pastures of the Mongolian landscape remain heavily degraded. A combination of high mobility and collective social support structures made Mongolia's semi-nomadic society highly adapted to fluctuating weather conditions and productivity of grasslands. However, climate and governance change has tested the resilience of these pastoralist communities. 
While results for donor-incentivized community-based resource management have been mixed, lessons can be learned from successful cases. Herd size regulations and agreements to share grazing lands among herder families could help sustain the livelihoods of families with smaller herd sizes \citep{fernandez2015}. When urbanization is the preferred adaptation/transformation strategy, governance mechanisms could be established that facilitate this transformation at an early stage, to avoid extended losses in utility (and wealth) resulting from continued adaptation to `flickering' environmental conditions. However, given already insufficient urban infrastructures, sprawling (and under-served) urban settlements and overwhelmed city management in cities around the world, including Mongolia, governance actors should carefully consider what types of interventions will contribute to greater sustainability of local human-environment systems. Importantly though, given the high cost of adaptation to flickering socio-environmental conditions, the high cost and resulting loss of utility associated with (mal-) adaptive efforts must be accounted for when weighing options for when and how to intervene. Governance interventions into rural and urban livelihoods may create new interdependencies and dynamics between social and ecological systems and between rural and urban environments.

\subsection*{Fisheries}
Fisheries support the income and food security of millions of people around the world \citep{mcclanahan2015managing}. Fisheries are also being severely impacted by climate change: from increases in ocean temperatures impacting the structural integrity of coral reefs and the fisheries they support \citep{Hoegh-Guldberg2017}, to ocean acidity impacting early life stages of fish \citep{waldbusser2015saturation} to species range shifts altering the spatial distribution of fishing effort \citep{pinsky2012lagged}, many fisheries are suffering, with subsequent adverse effects on coastal communities \citep{hollowed2013projected}. 

Fishers have developed several ways of dealing with these changes though. In particular, fishers can move where they go, tracking fish stocks as they shift with climate change \citep[although the price of fuel greatly constrains where there can go]{selden2020coupled}. Switching fisheries and operating in numerous fisheries over the course of a year is also common. This is one way in which fishers ``smooth'' their income over the year. But it can be costly in time and money, and sometimes impossible given certain fisheries management institutions (e.g., permits are required to fish, but sometimes not available). Switching between fisheries also requires training, knowledge and different fishing gear. 

Exposure to an ecological regime shift can restructure the incentives for what species are targeted. For example, coral reefs around the world are experiencing more frequent and intense marine heatwaves that lead to coral bleaching and mass coral mortality events \citep{Hoegh-Guldberg2017, Hughes2018}. In addition to impacts on coral cover, bleaching events are altering reef fish assemblages, especially if reefs experience a shift towards an algal-dominated state \citep{Richardson2018, Robinson2019}. Return times between bleaching events are presently about every six years and because coral requires on the order of 10-15 years for the fastest species to recover \citep{Gilmour2013}, it is possible that reefs today are  experiencing flickering between a high- vs. low-coral state or have already transitioned into the latter.  

For coral reef fisheries, fishers reduce their sensitivity to climate change, including the impacts of flickering between high- and low-coral cover states, through livelihood diversification \citep{cinner2012}. For example, in the Caribbean, alternative livelihoods among coastal fishers include agriculture, forestry, aquaculture, construction work, and ecotourism \citep{karlsson2020}. However, socio-economic barriers including poverty, a minimal social safety net, or a lack of access to capital can limit adaptive capacity and prevent this diversification~\citep{cinner2012}. Governance aimed at dismantling these barriers may also provide second-order benefits by helping to smooth the transition of socio-economic systems though the flickering stages of ecological regime shifts. The main concern suggested by our modeling is that if there is flickering between ecological/fishery states, then fishers might lose income by repeatedly adapting to different states. In addition to the costs associated with gaining the knowledge, the institutional costs such as attaining permits for a new fishery, and the sunk-costs associated with procuring necessary new fishing gear, may force many fishers to take drastic/transformative action, such as leaving fishing altogether. All of these factors indicate that a flickering transition in the underling ecosystem that a fishery is part of, is likely to cause a decline in wellbeing among fishers, potentially reducing the viability of certain fisheries in the future.

\subsection*{Forest management and wildfire risk}

Forest management can be viewed as embedded within a socio-environmental system. The ecological dynamics of forests and the stochastic dynamics of wildfire are both coupled with management practices including timber harvesting, fire suppression, and prescribed fire~\citep{luce2012climate, steelman2016}.

Climate change has led to increasing frequency and severity of drought, as well as hotter peak summer temperatures in the forest ecosystems of the Western US~\citep{mckenzie2004climatic}. These changes have coincided with increasing tree density in western forests which was driven, in part, by a long-term management emphasis on fire suppression~\citep{fellows2008}. Furthermore, there has been a dramatic increase in the extent of the wildland-urban interface~\citep{radeloff2018rapid}, which increases the likelihood of ignition events and elevates the magnitude of damages that could result from a wildfire. These increasing stressors, driven by climate change, management policies and development, have combined to increase wildfire risk~\citep{marlon2012} and raise the spectre of the collapse of these ecosystems and their transition to alternative states~\citep{adams2013mega}. Given inherent stochasticity in the ignition and spread of wildfires, forest ecosystems at the cusp of tipping points may exhibit flickering dynamics.

In response to these entangled and increasing risks, 
the USDA Forest Service has developed a wildfire crisis strategy  which centers on prescribed fire and other fuels treatments~\citep{wildfirecrisis2022}. However, current climatic conditions, high fuel loads, and the vast extent of the wildland-urban interface will make the transition towards fire-resilient landscapes challenging. 

Our model suggests that a transition to a fire-resilient landscape that exhibits flickering could strain people's wellbeing, especially when adaptive capacity of management agencies, communities and individuals is low. This results from the time it takes for social practices to align with environmental states. Further, a flickering transition to a fire-resilient landscape may contain periods of time that resemble the historical ecosystem state, where the pressure to invest in adaptations to increase fire resilience may seem unnecessary. Given flickering, these intercalary chapters of unpredictable duration which resemble historical conditions may end with dramatic shifts marked by wildfire. The Forest Service wildfire crisis strategy aims to minimize the risk of these conflagrations by pairing prescribed fire with mechanical fuels treatments so that fire intensity stays low and wildfire risk is diminished. This approach may decrease the risk of a flickering transition, but our modeling results suggest that a complimentary management focus on increasing people's adaptive capacity may blunt the negative impacts that a flickering transition can cause. For example, investments in programs which provide education about the risks of wildfire smoke~\citep{wen2022lower,burke2022exposures} and resources for improving indoor air quality could help communities navigate the transition to a fire resilient landscape with a decreased health burden from the impacts of air pollution.  

\section*{Discussion}
Global change is pushing socio-environmental systems to the brink of tipping points where further small changes to underlying conditions could lead to dramatic shifts in system states. Much work has focused on identifying early warning signals of tipping points and governance interventions that can prevent the collapse of the current socio-environmental state. However, as climate change and other anthropogenic impacts continue (e.g., converting Amazonian rainforests for agricultural and cattle use), we may see more social-ecological systems approach and pass through tipping points despite societies' best efforts to avoid them. In this case, in addition to efforts to avert a tipping point, it may be valuable to design governance interventions that minimize loss of wellbeing that results from  passing through a tipping point.  

Flickering, where a system switches among alternative stable states as a result of noise, can occur prior to a tipping point and serves as an early warning indicator~\citep{scheffer2009early}. However, the highly unpredictable environmental dynamics that occur under flickering pose unique challenges for people's environmental adaptation. In socio-environmental systems, rather than being an early warning signal, flickering may be a primary hurdle to successfully navigating a tipping point.
Policies that bolster people's adaptive capacity may be vital to ensuring means are available to adapt (and thrive) in coming decades. Tools such as (parametric) insurance~\citep{santos2021dynamics}, climate clubs~\citep{nordhaus2021dynamic} and risk pools~\citep{watson2018resilience,tilman2018revenue} are examples of mechanisms that could help people maintain adaptive capacity in the face of global change.  

A concerning possibility is that when people with low adaptive capacity are exposed to environmental flickering, the reduction in their wellbeing might further erode their adaptive capacity, for example by reducing their wealth or health. This could result in a vicious cycle wherein flickering induces a continuous reduction in the set of adaptation options open to people. Such a cycle could induce conditions that force people into adopting outside options, including urban or international migration. Human migration driven by adverse environmental conditions is a well known consequence of climate change~\citep{cattaneo2020human}. A term used in the context of sea-level rise to describe the inevitability of community reorganization in the face of environmental change is ``managed retreat''~\citep{alexander2012managed}. The challenge then is to ensure that this retreat is indeed managed in a way that accounts for the unique impacts that flickering could have during a transition through a regime shift. For many people around the world some form of retreat might be inevitable, either in space (i.e., human migration) or in terms of job sector.

Income diversification is among the primary tools that people use for dealing with environmental risks~\citep{brouwer2007socioeconomic,shah2021livelihood}. Therefore, flickering could incentivize people to diversify or change the industries that secure their income. This is evident in fisheries, where working in numerous fisheries throughout the year can act as a natural means of buffering the stochasticity associated with harvest~\citep{kasperski2013income,finkbeiner2015role,cline2017fisheries}. Fishers are also known to work in multiple sectors, including jobs on land within the timber and agricultural sectors~\citep{anderson2017benefits}. Environmental flickering in the oceans could lead to fishers redistributing their efforts over the set of fisheries available to them, and other industries that they work in. Analogously to migration, extreme environmental flickering might induce people to permanently leave one industry for another. 

While we have studied the impact passing through a single regime shift on people's wellbeing, in reality there may be multiple cascading environmental tipping points~\citep{rocha2018cascading}. In this case of systemic environmental risk, the associated social and environmental flickering could co-occur across various dimensions of a person's income portfolio. This might mean fishers will be unable to adapt by moving sectors, and nomadic herders may no longer find community support structures to help them recover from devastating losses. These impacts could scale-up and result in global systemic risks~\citep{centeno2015emergence} due to the connected nature of our environment, our socio-technological, and our governance systems. Correlated risks among marine heatwaves at sea, droughts on land and economic volatility could interact to present large-scale challenges for communities. 

Early-warning signals of environmental regime shifts may help people manage their adaptation to impending change~\citep{lenton2011early}. However, we find that some early-warning signals of environmental regime shifts describe socio-environmental dynamics that can already have negative impacts on people. In these cases, our results suggest three types of governance interventions may be warranted either individually, or in combination. First, investments in assuring that people have high adaptive capacity can mitigate the impacts of flickering on wellbeing by helping people remain well adapted to the rapidly changing environment. Second, facilitating transformational change, which partially decouples environmental adaptation and wellbeing, can result in greater wellbeing across a broad range of conditions. Lastly, when a wellbeing trough is unavoidable, interventions that facilitate people's transitions to different ways of life or migration to different places may be warranted.  
Critically, these investments appear to be most benefical if enacted well before a tipping point is crossed. 
This suggests that climate adaptation policies may need to be more anticipatory than originally thought.

\section*{Acknowledgements}
The findings and conclusions in this publication are those of the authors and should not be construed to represent any official USDA or U.S. Government determination or policy. Simulation code is available at \href{https://github.com/atilman/RegimeShift_Wellbeing}{https://github.com/atilman/regimeshift\_wellbeing}.

\bibliographystyle{apalike}
\bibliography{refs}

\appendix
\renewcommand{\thesection}{S.I. \arabic{section}}    %%%% but here
\newpage
\begin{center}
\large\textit{Supplementary Information for:}
\end{center}\vspace{-30pt}
\section*{Maintaining human wellbeing as socio-environmental\\ systems undergo regime shifts}
\renewcommand{\thefigure}{SI~\arabic{figure}}
\setcounter{figure}{0}
\renewcommand{\thetable}{SI~\arabic{table}}
\setcounter{table}{0}

\begin{figure}[ht]
    \centering
    \includegraphics[width = .7\textwidth]{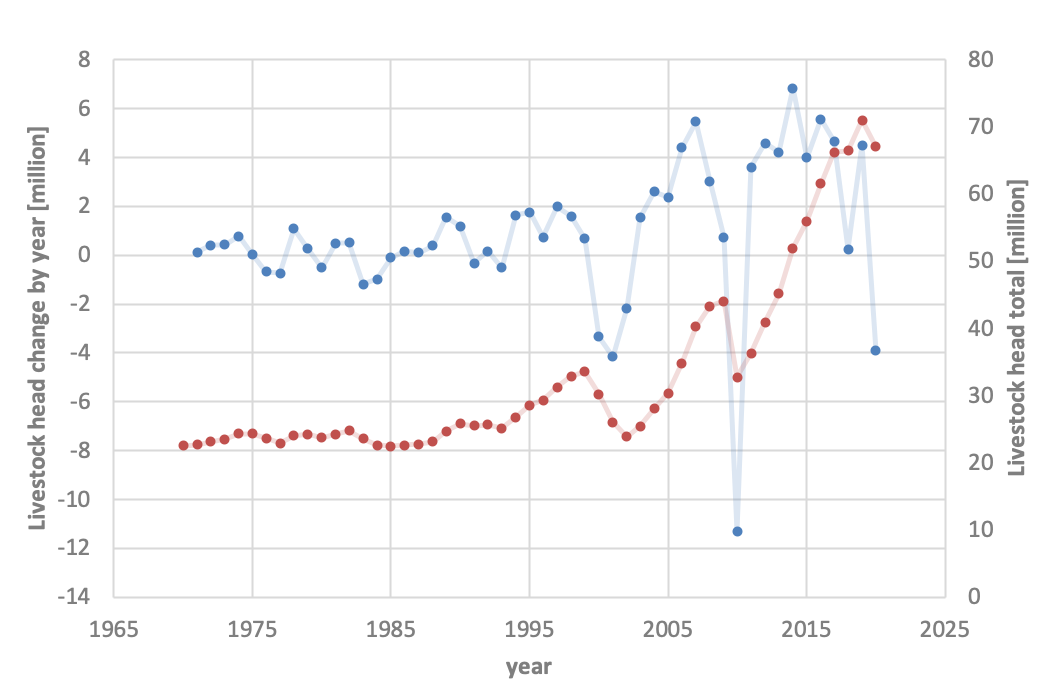}
    \caption{Development of livestock numbers 1966-2020. Restrictions on herd sizes and total head count were lifted after the transition from the socialist to the capitalist regime in 1990. The significant growth in livestock numbers starting in the mid-1990's is followed by strong dynamics resulting from large livestock losses and recoveries due to climatic extremes and changing adaptive strategies. The growth in livestock numbers coincides with a drop in NDVI and rising potential evapotranspiration rates recorded by \citep{Enebish2020}. Data publicly available from \cite{MongoliaStats}.}
    \label{SI fig:grazing rate}
\end{figure}

\begin{figure}
    \centering
    \includegraphics[width = .7\textwidth]{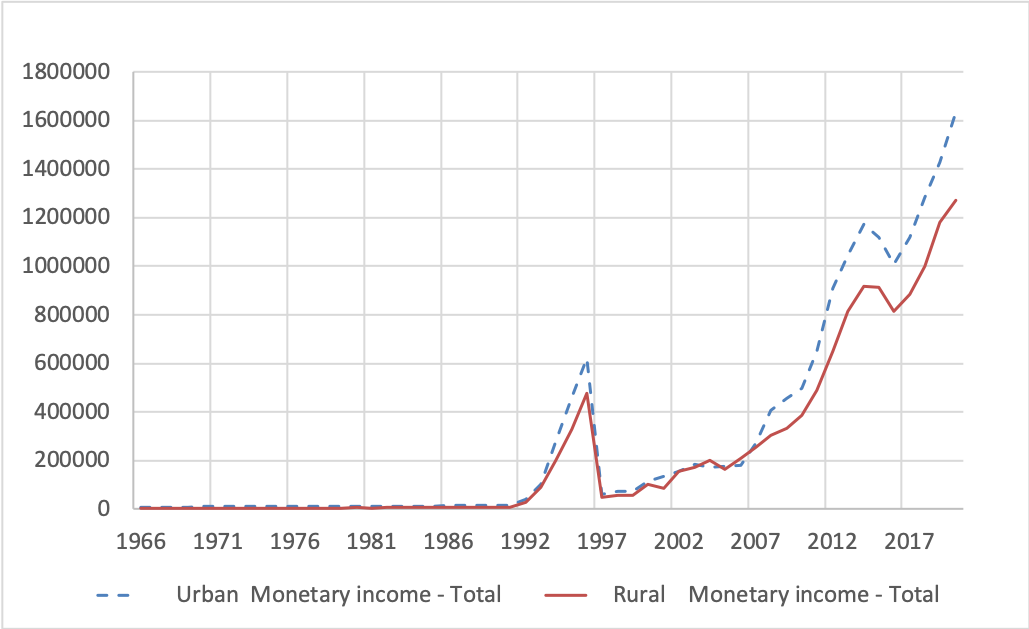}
    \caption{Development of monetary income in Mongolian Tugrik. The drop in the mid-1990's reflects the impacts of the 1997 Asian financial crisis. A gap emerges between urban and rural incomes from the late 2000's. Data publicly available from \cite{MongoliaStats}.}
    \label{S2 Income}
\end{figure}

\begin{figure}
    \centering
    \includegraphics[width = .7\textwidth]{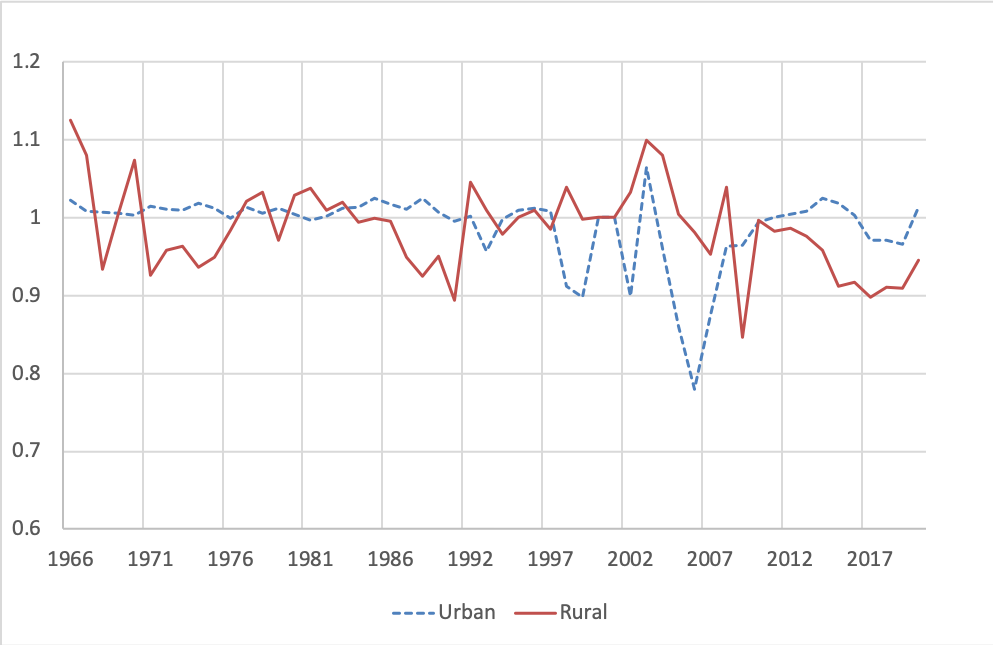}
    \caption{Income/expenditure ratio for rural and urban households 1966-2020. Urban household income/expenditure ratios remain relatively steady until the 1997 and continue to fluctuate strongly until ~2010, when they become more steady again. This could indicate the transition to an urban regime, as urbanization started to increase significantly in the late 1990's. Rural income/expenditure ratios are more dynamic from year-to-year, and while they have followed urban dynamics more closely since 2010, there has been a large gap between urban and rural, with rural income/expenditure ratios being significantly lower that urban ratios. This indicates the costly toll on average utility resulting from the high costs of adaptation to the flickering regime for rural households. Data publicly available from \cite{MongoliaStats}.}
    \label{S3 Income/Expenditure}
\end{figure}

\begin{figure}
    \centering
    \includegraphics[width = .7\textwidth]{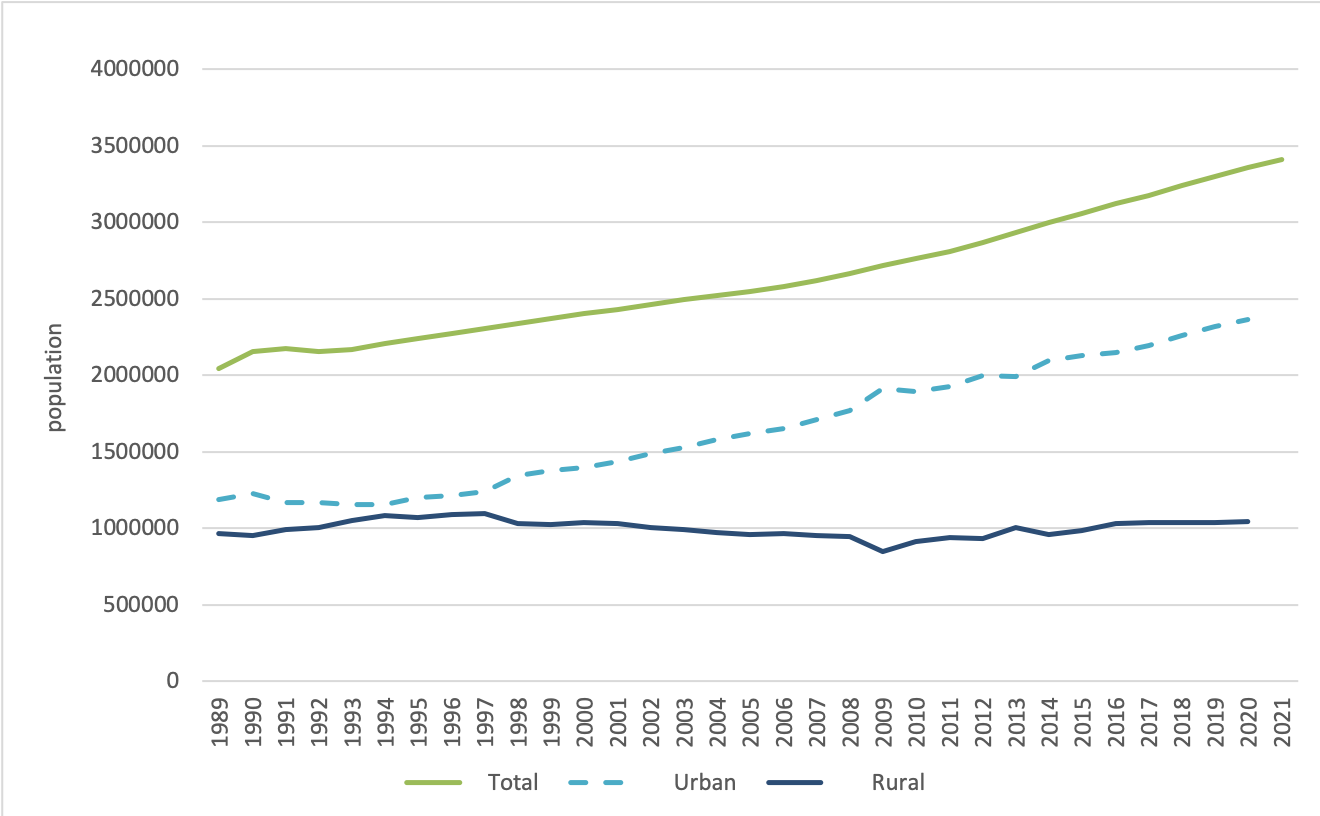}
    \caption{Population development in Mongolia 1989-2021. Data publicly available from \cite{MongoliaStats}.}
    \label{SI fig: Population}
\end{figure}

\begin{figure}
    \centering
    \includegraphics[width = .7\textwidth]{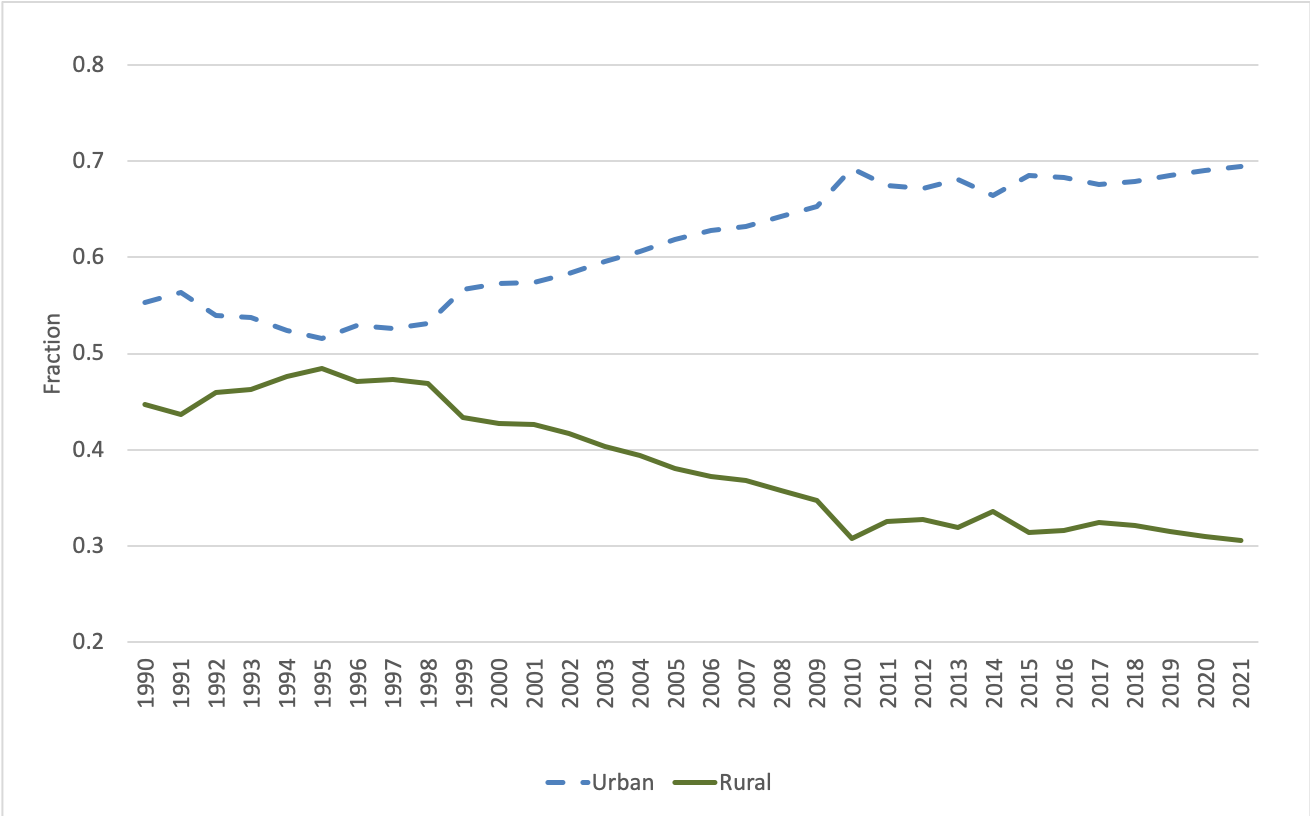}
    \caption{Fraction of population living in urban and rural areas in Mongolia 1989-2021. Data publicly available from \cite{MongoliaStats}.}
    \label{SI fig: PopulationUrban}
\end{figure}
\begin{landscape}
\begin{table}[ht]
\begin{center}
    \begin{tabular}{c||c|c|c|c|c|c } 
     Parameter & Fig. 2  &Fig. 3  & Fig. 4 & Fig. 5 & Fig. 6& Description \\ 
     \hline\hline
     $l$ & n/a& n/a & 0.01 & 0.001; 0.01; 0.1& .001& Adaptation rate\\
     \hline
     $r$ & 1& n/a & 1 & 1& 1& Resource growth rate\\
     \hline
     $K$ & 10& n/a & 10 & 10& 10& Carrying capacity\\
     \hline
     $h$ &1& n/a & 1 & 1& 1& Extraction half-saturation constant\\
     \hline
     $\mu$ & n/a& n/a & 0 & 0& 0& Mean noise\\
     \hline
     $\beta$ & n/a& n/a & 0.07 & 0.07& 0.07& Noise standard deviation\\
    \hline
     $T$ & n/a& n/a & 30 & 30& 30& Timescale of noise decorrelation\\
          \hline
     $a$ & n/a& 3; 5 & 3 & 3& 3; 5& Misadaptaiton at which utility is halved\\
          \hline
     $m$ & n/a& 5; 5.75& 5 & 5& 5; 5.75& Payoff function intercept\\
          \hline
     $n$ & n/a& 1/2; 1/10 & 1/2 & 1/2 & 1/2; 1/10& Payoff function slope\\
          \hline
     $c$ & 1-3.5& n/a & 1; 1.95; 2.45; 3.1 & 0.25-3.5& 0.25-3.5& Resource extraction rate\\
    \end{tabular}
     \caption {Parameter values used to generate each main text figure.}
     \label{SI Table: paramVals}
\end{center}
\end{table}
\end{landscape}
\end{document}